\documentclass{appolb}
\usepackage{graphicx}
\usepackage{lineno}

\begin{document}
\title{Overview of recent femtoscopy measurements with ALICE
\thanks{Presented at XIII Workshop on Particle Correlations and Femtoscopy, Cracow, Poland, 2018.}
}
\author{Ma{\l}gorzata Anna Janik (for the ALICE Collaboration)
\address{Warsaw University of Technology, ul. Koszykowa 75, 00-662 Warsaw, Poland}
}

\maketitle
\vspace{-0.3cm}
\begin{abstract}
One of the key methods used in the study of the Quark-Gluon Plasma (QGP) is femtoscopy, the technique of measuring short-range two-particle correlations as a function of relative momentum. Traditionally, femtoscopy has been utilized to measure the size of the QGP fireball created in relativistic heavy-ion collisions. However, since it is sensitive to the correlations between the particles in the final state, it has been shown recently that the parameters of the strong interaction can be probed as well. 
This review includes a broad range of ALICE femtoscopy results, including both traditional and novel measurements of the interaction between particles.

\end{abstract}
\vspace{-0.3cm}
\PACS{13.60.Le, 13.60.Rj, 13.75.Cs, 25.75.−q}

\vspace{-0.3cm}
\section{Introduction}
The ALICE detector at the Large Hadron Collider (LHC) at CERN has been optimized for excellent particle identification and tracking with the aim of studying the Quark-Gluon Plasma (QGP), a deconfined state of hadronic matter produced in ultra-relativistic heavy-ion collisions at the LHC. Utilizing the detector's unique capabilities allows one to perform a broad range of femtoscopic measurements -- studies of particle correlations in relative momentum space. In this proceedings some of the latest femtoscopy results from ALICE are reported.

\vspace{-0.3cm}
\section{Traditional femtoscopy}

The femtoscopic technique makes it possible to measure space-time characteristics of the particle-emitting source using particle correlations in momentum space. The Koonin-Pratt equation \cite{Koonin:1977fh,Pratt:1990zq}, which relates the experimental correlation function $C(k^{*})$  with the source emission function $S(\vec{r}^{*})$, is given by
\begin{equation}
C(k^{*})=\int S(\vec{r}^{*}) \left| \Psi_{-\vec{k}^{*}}(\vec{r}^{*}) \right|d^3\vec{r}^{*}.
\label{eq:femtodef}
\end{equation}
The source function is usually assumed to be a Gaussian $S(\vec{r}^{*}) \sim e^{-\left| \vec{r}^{*} \right| ^2/(4R^2)}$ ($R$ is the observable corresponding to the source size, also referred to as the ``HBT radius''), while $\left| \vec{r}^{*} \right|$ is the relative distance in the pair-rest frame and $\Psi_{-\vec{k}^{*}}(\vec{r}^{*})$ is the two-particle interaction kernel. Depending on the studied pair $\Psi$ may include effects arising due to quantum statistics and/or final-state interactions.

In ``traditional femtoscopy" we are interested in the emission function  $S(\vec{r}^{*})$ -- we want to study the produced system and determine its size and shape. We use our knowledge of the interaction kernel $\Psi$ and the measurement of $C(k^{*})$ to study the characteristics of the source.

\subsection{Kaon-kaon and pion-kaon femtoscopy}
\label{sec:kaons}

The most commonly studied particles in femtoscopic measurements are pions, due to their abundance. However, studies of kaons offer cleaner signals (less affected by resonances), and models which describe pions well should describe kaons with equal precision. Moreover, comparison of pions and kaons as a function of $m_{\rm T}$ could give us information about the rescattering phase effects -- the rescattering phase should have a different influence on pions and kaons, which in turn can lead to broken $m_{\rm T}$-scaling. 

In \cite{Acharya:2017qtq} three-dimensional kaon femtoscopy analysis of Pb-Pb collisions at $\sqrt{s_{\rm NN}}=2.76$~TeV is reported, using both neutral and charged kaons.  Data is compared to models with and without a rescattering phase. Broken $m_{\rm T}$-scaling indicates the importance of the hadronic rescattering phase at LHC  energies. Additionally, emission times for pions and kaons are extracted. The measured emission time of kaons is larger than that of pions by $\approx2.1$~fm/$c$.

\begin{figure}[!ht]
	\centering
	\includegraphics[width=7.1cm,keepaspectratio]{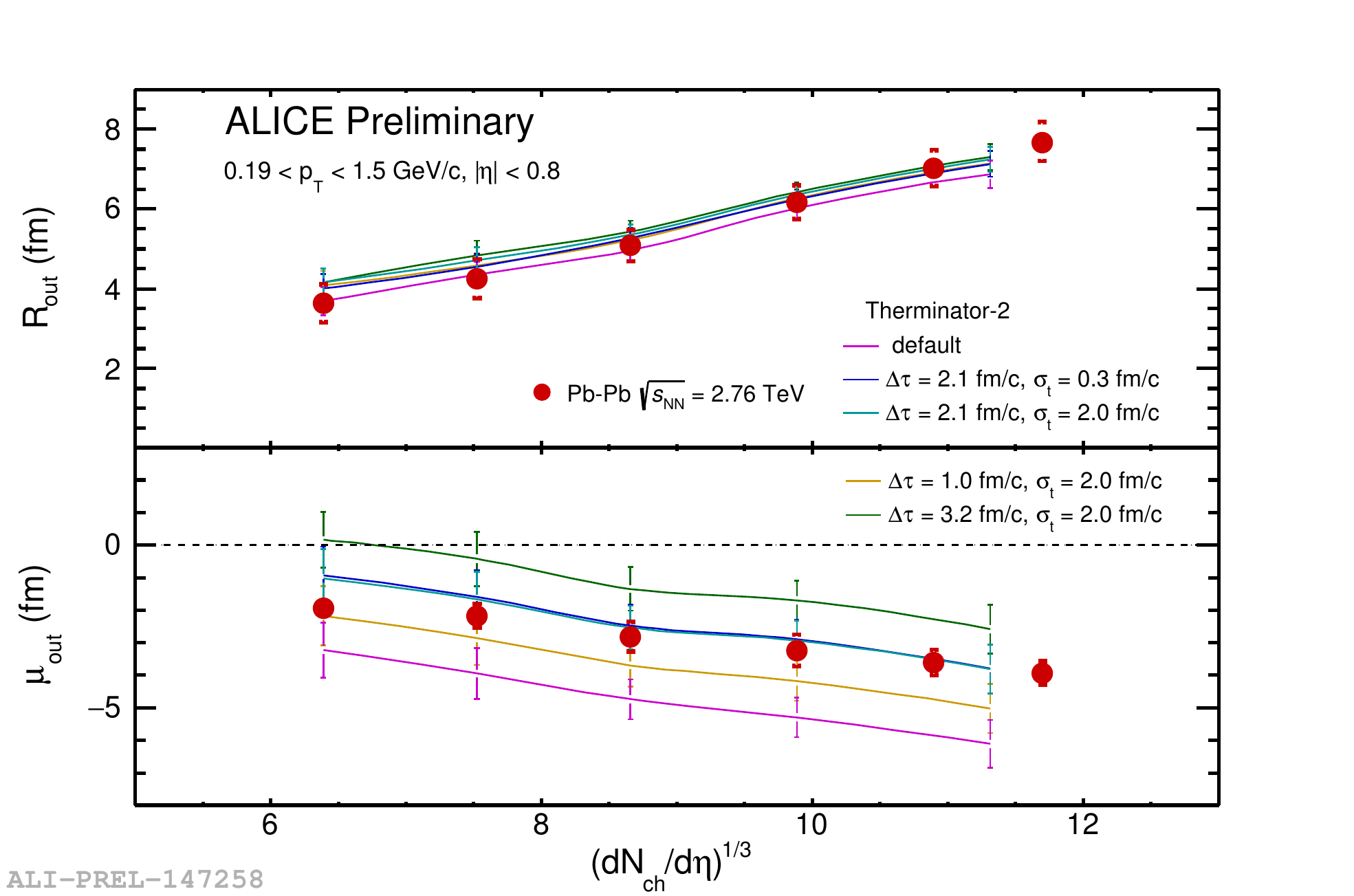}
	\includegraphics[width=4.9cm,keepaspectratio]{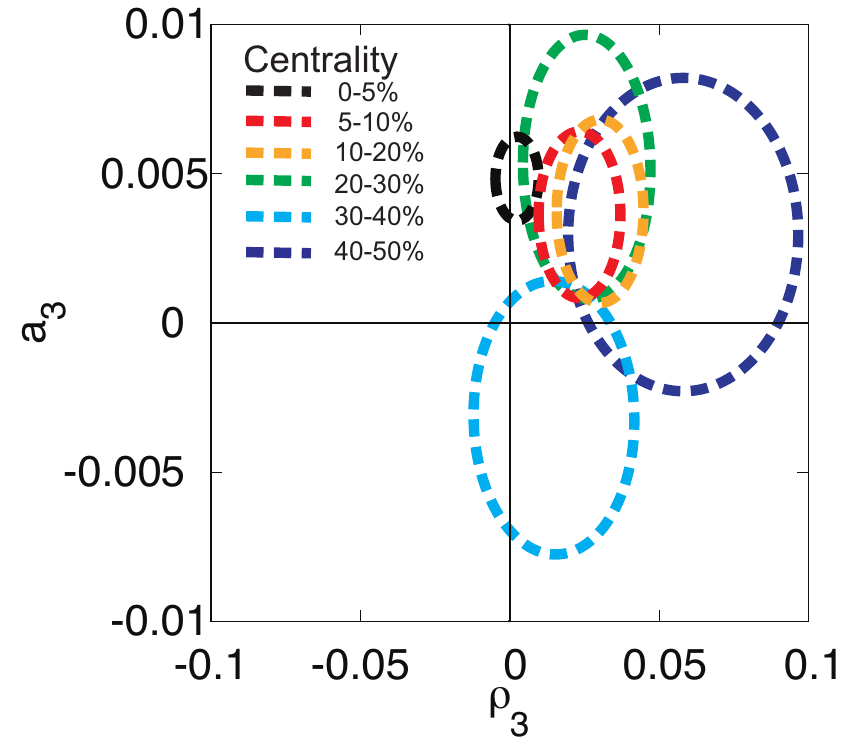}\\
	\caption{Left: Source size (top panel) and pion--kaon emission asymmetry (bottom panel) from pion--kaon correlation functions. The solid lines show predictions from a calculation of source-size and emission asymmetry using Therminator-2 model. Plot from \cite{Pandey:2018vrp}. Right: Blast-Wave model source parameters, final source anisotropy ($a_3$) and transverse flow ($\rho_3$), for different centrality ranges, as obtained from the fit to ALICE radii oscillation parameters. The contours represent the one sigma uncertainty. Plot from \cite{Acharya:2018dpu}.
	}
	\label{fig:AshuthoshAndAzimuthally}
\end{figure}

Similar studies of non-identical particle pairs, like pion-kaon correlations, can give us more information about the space-time asymmetry between studied particles.  In \cite{Pandey:2018vrp} it is shown that  the average source size of the system and the emission asymmetry between pions and kaons increase from peripheral to central events (see also Fig.~\ref{fig:AshuthoshAndAzimuthally}-left). Asymmetry is significantly negative, which  means that on average pions are emitted closer to the centre and/or later than kaons. Additional model studies confirm that the pion-kaon data is consistent with  $\approx2.1$~fm/$c$ delay seen by pion-pion and kaon-kaon analysis. This confirms the observation that different particle species freeze-out at different times.

\vspace{-0.2cm}
\subsection{Azimuthally differential femtoscopy}
\label{sec:azimuthally}

Azimuthally differential femtoscopy allows us to study the source size
relative to the reaction plane.
In  \cite{Adamova:2017opl} the first azimuthally differential measurements of the pion source size relative to the second harmonic event
plane in Pb-Pb collisions at $\sqrt{s_{\rm NN}}=2.76$~TeV were reported. This observable gives us information about spatial geometry. The data shows that the pion source at the freeze-out is elongated in the out-of-plane direction, similar to what was observed at RHIC \cite{Adare:2014vax,Adamczyk:2014mxp}.

Moreover, pion source size relative to the third harmonic event plane  in Pb-Pb collisions at $\sqrt{s_{\rm NN}}=2.76$~TeV was studied as well, see \cite{Acharya:2018dpu}. HBT radii oscillations relative to the $3^{\rm rd}$ harmonic event plane are predominantly defined by the velocity fields; therefore, radii oscillations observed in data signal a collective expansion and anisotropy in these fields. A comparison of the measured radii
oscillations with the Blast-Wave model calculations indicate that the initial state triangularity is washed-out at freeze out (final source anisotropy is close to zero, see Fig.~\ref{fig:AshuthoshAndAzimuthally}-right).

\vspace{-0.2cm}
\section{Beyond the system size}
Apart from the studies of the space-time characteristics of the source, we can also use femtoscopic measurements to study pair wave function that includes information about interaction cross sections. For the pairs of particles for which the interaction is poorly or not known we can use other measurements to fix the source function $S(\vec{r}^{*})$ and use Eq.~(\ref{eq:femtodef}) to access information about $\Psi$.
The pair wave function $\Psi$ can be parameterized with a scattering length $f_0$ and effective radius $d_0$ parameters. The correlation function is then characterized by three parameters: $R$, $f_0$ and $d_0$.

\begin{figure}[!ht]
	\centering
	\includegraphics[width=9cm,keepaspectratio]{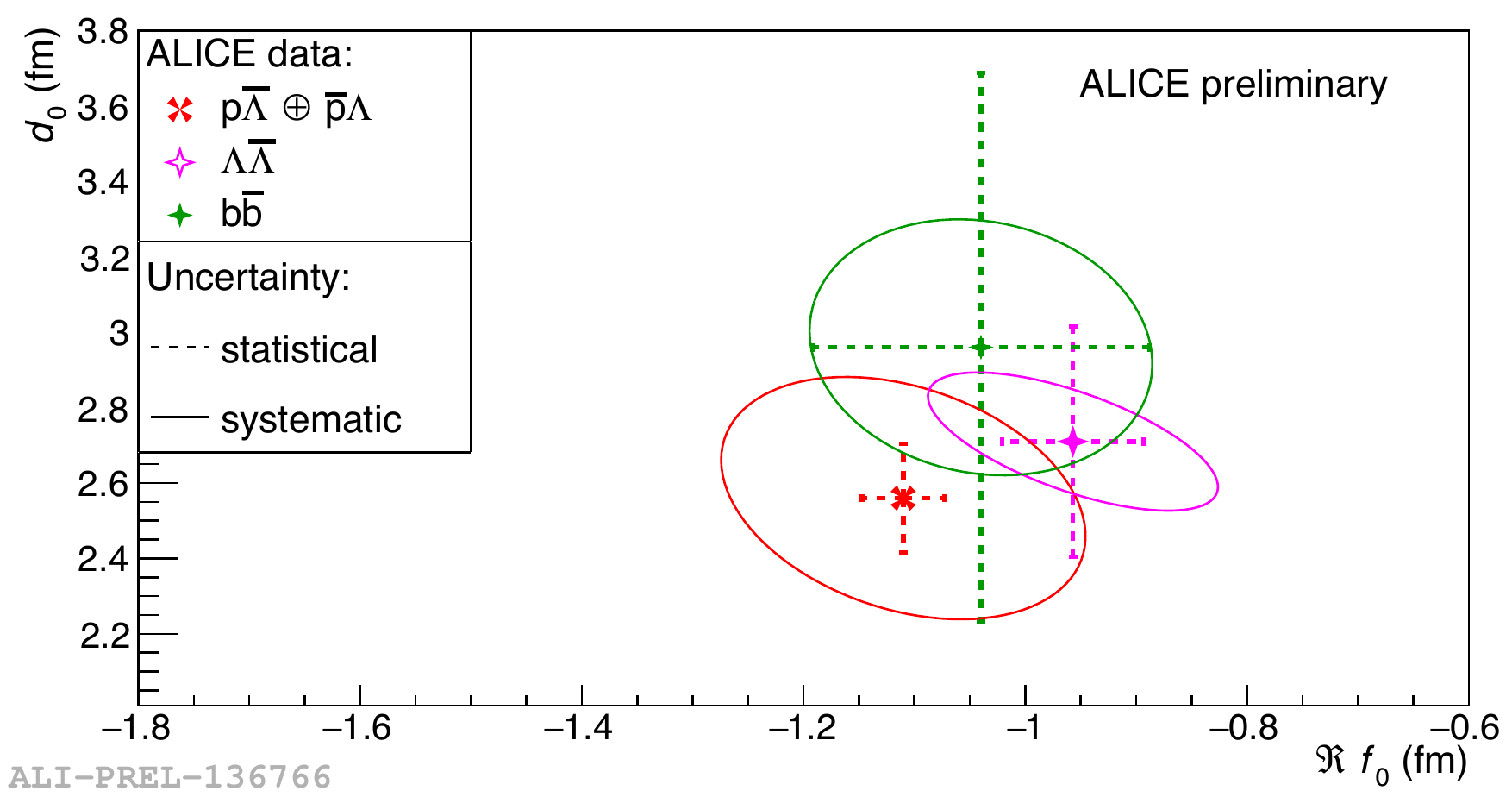}
	\caption{Strong interaction parameters for different baryon-antibaryon pairs. Plot from \cite{Niedziela:2018fau}.
	}
	\label{fig:Jeremi}
\end{figure}

ALICE has studied interaction parameters between a number of particle pairs. In \cite{Niedziela:2018fau} baryon-antibaryon correlations were studied using a ``global" fit to a large set of functions: 2 energies (both  $\sqrt{s_{\rm NN}}=2.76$~TeV and  $\sqrt{s_{\rm NN}}=5.02$~TeV), 3 pair combinations ($p\bar{p}$, $p\bar{\Lambda}+\bar{p}\Lambda$, and $\Lambda\bar{\Lambda}$), and in 6
centrality ranges. Results (see Fig.~\ref{fig:Jeremi}) show that scattering parameters for all baryon-antibaryon pairs are similar to each other. Moreover, we observe a negative real part of the scattering length which can indicate repulsive
strong interaction or creation of a bound state. Significant positive imaginary part of scattering length is a manifestation of a non-elastic
channel -- annihilation.

\begin{figure}[!ht]
	\centering
    \includegraphics[width=11.5cm,keepaspectratio]{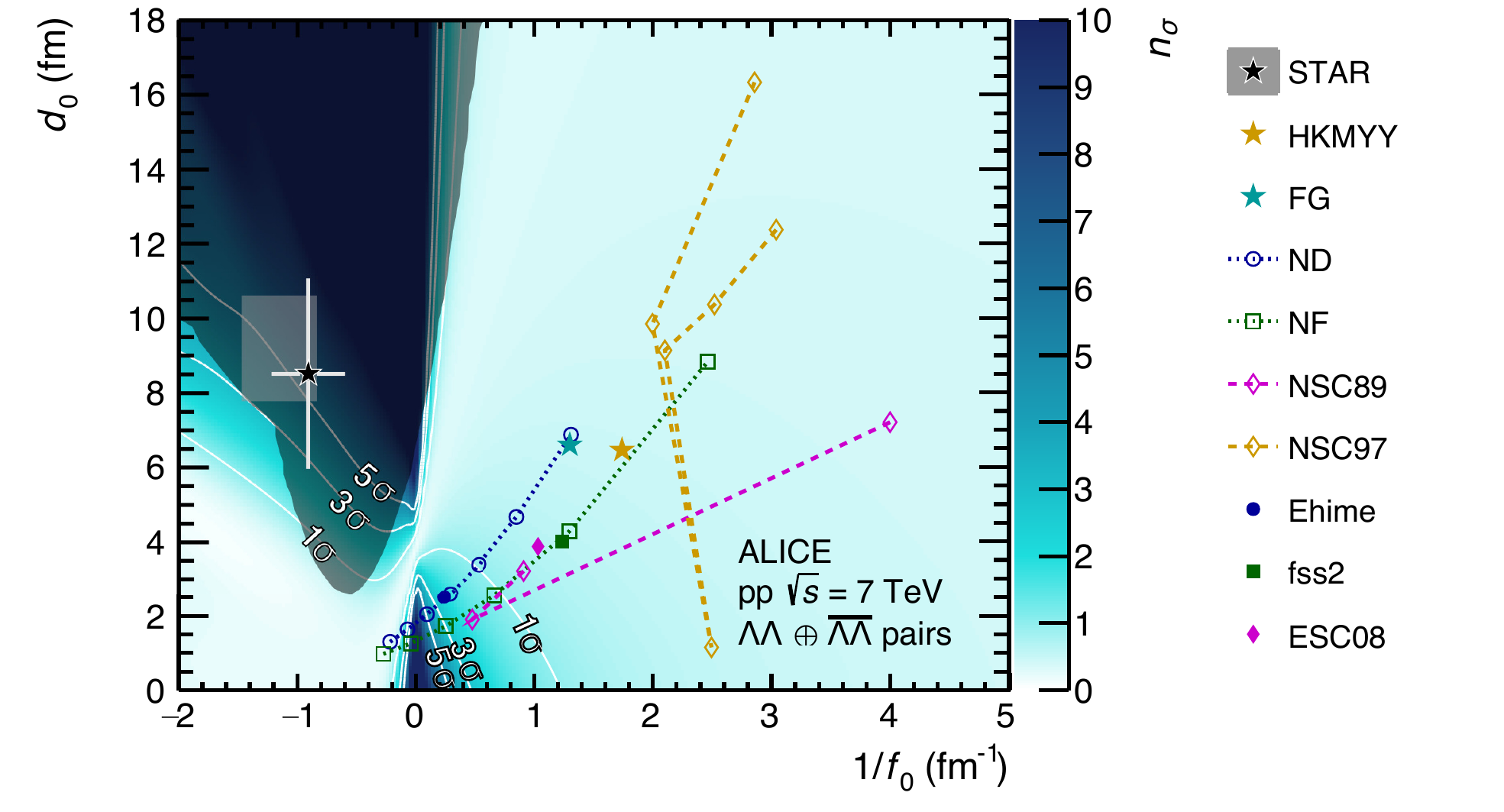}
	\includegraphics[width=11.5cm,keepaspectratio]{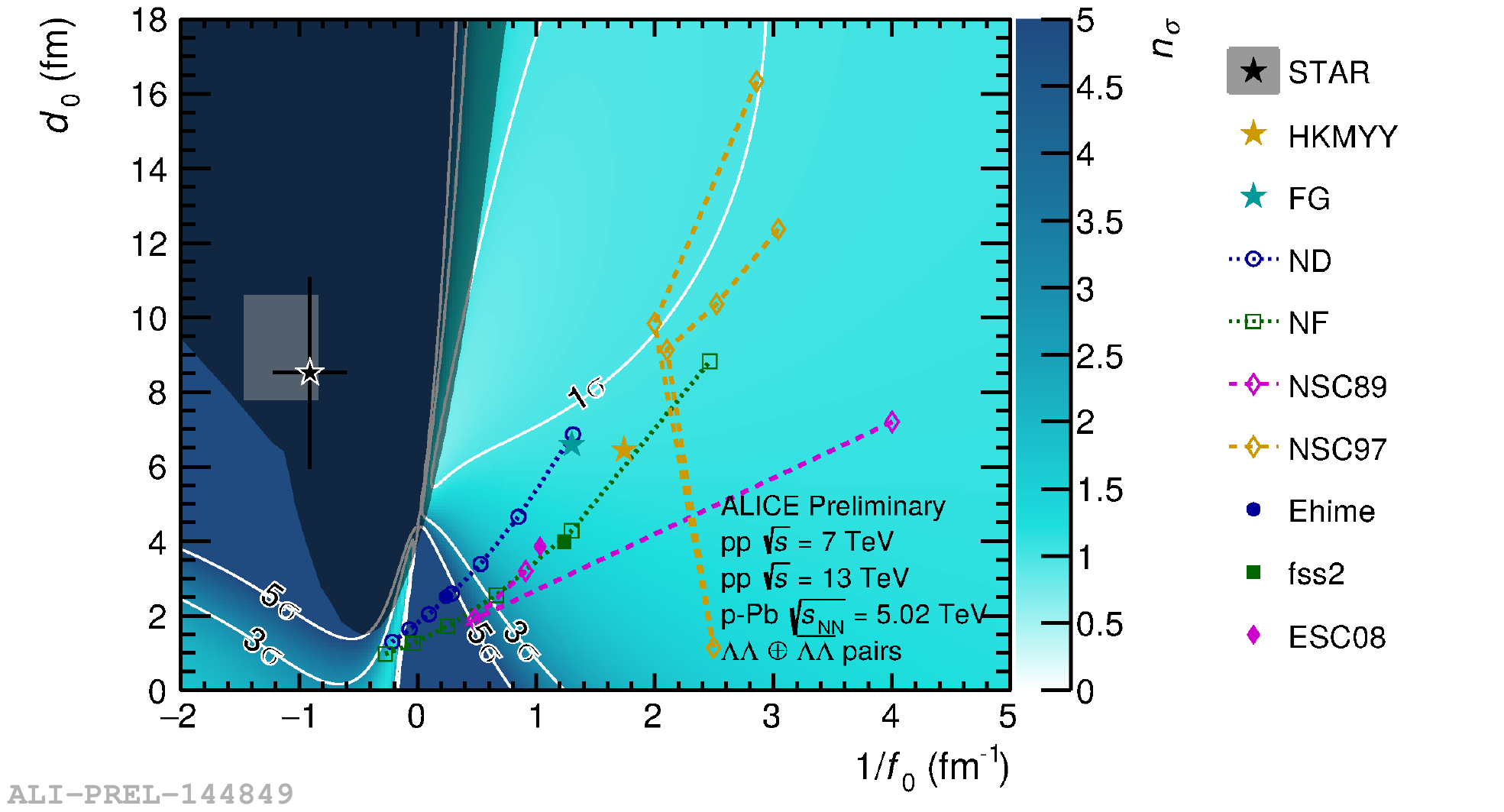}\\
	\caption{Number of standard deviations of the modeled correlation function for a given set of scattering parameters (effective range $d_0$ and scattering length $f_0$) with respect to the data, together with various model calculations (points and lines in colors) and STAR measurement (black star). Left: pp collisions at $\sqrt{s}= 7$~TeV. From \cite{Acharya:2018gyz}. Right: pp collisions at $\sqrt{s}= 7$ and $13$~TeV and p--Pb collisions at $\sqrt{s_{\rm NN}}= 5.02$~TeV. From \cite{Bernard}.
	\vspace{-0.5cm}}
	\label{fig:Laura}
\end{figure}

Except for pairs like proton-proton or proton-neutron, cross sections
for other baryon-baryon pairs are practically unknown. In \cite{Acharya:2018gyz} the femtoscopy method is employed to constrain the hyperon--nucleon and hyperon--hyperon interactions in pp collisions at $\sqrt{s}=7$~TeV. Results demonstrate that such studies are feasible in small systems, considering no minijet background is observed for the correlation functions of baryons (see also \cite{Adam:2016iwf}). The present data in the $\Lambda\Lambda$ channel, see Fig.~\ref{fig:Laura}-left, allows us to constrain the available scattering parameter space. 
Ref.~\cite{Bernard} and Fig.~\ref{fig:Laura}-right show the $\Lambda\Lambda$ scattering parameters constrained by all the $pp$ and p--Pb collision data recorded thus far during the LHC Run 2, almost excluding negative values of $f_0$.
We expect up to 100 times more data in the continuation of Run 2 as well as Run 3 at the LHC, which will enable us to measure these parameters with greater precision and extend the method to other baryons.

\section{Summary}
ALICE has provided a wealth of femtoscopy measurements. Studies of pions and kaons suggest that kaons are emitted later than pions. The data show a $2.1$ fm$/c$
delay between pion and kaon average emission times. The azimuthally differential analysis shows that the source at the freeze-out is elongated in the out-of-plane direction. The initial state triangularity is washed-out at freeze out.
Moreover, with this new technique ALICE can probe strong interaction cross sections which cannot be currently measured in any other way. This opens a whole new area of studies that ALICE plans to pursue and expand in the future.


\vspace{-0.2cm}
\section*{Acknowledgements}
\vspace{-0.1cm}
This work was supported by the Polish National Science Centre under decisions no. UMO-2015/19/D/ST2/01600, no. UMO-2016/22/M/ST2/00176 and no. UMO-2017/27/B/ST2/01947.
\vspace{-0.3cm}
\bibliographystyle{ieeetr}

\end{document}